\documentclass[12]{article}
\usepackage{fleqn}
\DeclareMathAlphabet{\mathpzc}{OT1}{pzc}{m}{it}
\usepackage{amsmath}
\usepackage{graphicx}
\begin{document}

\Large
\begin{center}
\bf{Magic Three-Qubit Veldkamp Line and\\ Veldkamp Space of the Doily}
\end{center}
\vspace*{-.3cm}

\large
\begin{center}
 Metod Saniga$^{1}$ and Zsolt Szab\'o$^{2}$
\end{center}
\vspace*{-.5cm} 

\normalsize
\begin{center}

$^{1}$Astronomical Institute of the Slovak Academy of Sciences,\\
SK-05960 Tatransk\' a Lomnica, Slovak Republic\\
(msaniga@astro.sk)

and

$^{2}$Department of Theoretical Physics, Institute of Physics,\\ Budapest University of Technology and Economics, H-1521 Budapest, Hungary\\ (zs20002@gmail.com)

\end{center}

\vspace*{-.4cm} \noindent \hrulefill

\vspace*{-.1cm} \noindent {\bf Abstract}

\noindent
A magic three-qubit Veldkamp line of $W(5,2)$, i.\,e. the line comprising a hyperbolic quadric $\mathcal{Q}^+(5,2)$, an elliptic quadric $\mathcal{Q}^-(5,2)$ and a quadratic cone $\widehat{\mathcal{Q}}(4,2)$ that share a parabolic quadric $\mathcal{Q}(4,2)$, the doily, is shown to provide an interesting model for the Veldkamp space of the latter. The model is based on the facts that: a) the 20 off-doily points of $\mathcal{Q}^+(5,2)$ form ten complementary pairs, each corresponding to a unique grid of the doily; b) the 12 off-doily points of $\mathcal{Q}^-(5,2)$ form six complementary pairs, each corresponding to a unique ovoid of the doily; and  c) the 15 off-doily points of $\widehat{\mathcal{Q}}(4,2)$  -- disregarding the nucleus of $\mathcal{Q}(4,2)$ -- are in bijection with the 15 perp-sets of the doily.
These findings lead to a conjecture that also parapolar spaces can be relevant for quantum information.
\\

\vspace*{-.2cm}


\vspace*{-.2cm} \noindent \hrulefill

\bigskip
\noindent
\section{Introduction}
Quantum information theory (QIT), an important  branch of quantum physics, is the study of how to integrate information theory with quantum mechanics, by studying how information can be stored in (and/or retrieved from) a quantum mechanical system. Within the last ten to fifteen years it has been gradually realized  that finite geometries represent key mathematical concepts of QIT.
Among them, the {\it unique} triangle-{\it free} $15_3$-configuration (out of 245,342 ones), also known as the Cremona-Richmond configuration and in the sequel referred to as the doily, 
acquires a special footing. This notable role of the doily stems from the fact that it is isomorphic to three remarkable, conceptually-distinct point-line incidence structures, namely 
a symplectic polar space of type $W(3,2)$ (whose subgeometries furnish simplest observable proofs  of quantum contextuality and justify the existence of the maximal sets of MUBs in the associated Hilbert space of two-qubits \cite{plasa}), an orthogonal parabolic polar space of type $\mathcal{Q}(4,2)$ (being the core  of the magic three-qubit Veldkamp line of form theories of gravity \cite{lhs}) and a generalized quadrangle of type GQ$(2,2)$ (being a subquadrangle of GQ$(2,4) \cong \mathcal{Q}^-(5,2)$ that entails some important aspects of the so-called black-hole/qubit correspondence \cite{lsvp}). Employing the concept of Veldkamp space of a point-line incidence structure, this note aims at shedding some interesting light on how these three geometrical settings are interrelated.

\section{Basic Glossary}
We shall start with a well-known duad-syntheme model of the doily \cite{paythas}.
Given a six-element set  $S=\{1,2,3,4,5,6\}$, let us call a two-element subset of $S$ a duad, and a set of three duads forming a partition of $S$ a syntheme. The point-line incidence structure whose points are ${6 \choose 2} = 15$ duads and whose lines are ${6 \choose 2}{4 \choose 2}{2 \choose 2}/
3! = 15$ synthemes, with incidence being containment, is isomorphic to the doily. 

A generalized $n$-gon $\mathcal{G}$; $n \geq 2$, is a point-line incidence geometry which satisfies the following two axioms \cite{mal}:
a) $\mathcal{G}$ does not contain any ordinary $k$-gons for $2 \leq k < n$ and b)  given two points, two lines, or a point and a line, there is at least one ordinary $n$-gon in $\mathcal{G}$ that contains both objects.
A finite generalized $n$-gon $\mathcal{G}$ is of order $(s, t), s, t \geq 1$, if every line contains $s+1$ points and 
every point is contained in $t + 1$ lines; if $s = t$, we also say that $\mathcal{G}$ is of order $s$.
A generalized $4$-gon is also called a {\it generalized quadrangle} (and abbreviated as GQ).

Next, we will also need particular types of four different kinds of finite polar spaces \cite{cam}.
The {\it symplectic} polar space $W(2N - 1,q)$, $N \geq 1$, 
consisting of all the points of PG$(2N - 1, q)$ together with the totally isotropic subspaces in respect to the standard symplectic form 
$$\theta(x,y) = x_1 y_2 - x_2 y_1 + \dots + x_{2N-1} y_{2N} - x_{2N} y_{2N-1}.$$
The {\it hyperbolic} orthogonal polar space $\mathcal{Q}^{+}(2N - 1, q)$, $N \geq 1$, 
formed by all the subspaces of  PG$(2N - 1, q)$ that lie on a given nonsingular hyperbolic quadric, with the standard equation $$x_1 x_2 + x_3x_4 + \ldots + x_{2N-1} x_{2N} = 0.$$
The {\it elliptic} orthogonal polar space $\mathcal{Q}^{-}(2N + 1,q)$, $N \geq 1$,
featuring all points and subspaces of PG$(2N + 1, q)$ satisfying the standard equation $$f(x_1,x_2)+x_3x_4+\cdots+x_{2N+1}x_{2N+2} = 0,$$ where $f$ is irreducible over $GF(q)$.
And, finally, the {\it parabolic} orthogonal polar space $\mathcal{Q}(2N,q)$, $N \geq 1$,
formed by all points and subspaces of PG$(2N, q)$ satisfying the standard equation $$x_1x_2 + x_3x_4 + \cdots + x_{2N-1}x_{2N} + x_{2N+1}^2 = 0.$$ 
A projective subspace of maximal dimension is called a generator; all generators have the same (vector) dimension $r$, which is called the rank of the polar space.
The rank of each of the above-listed cases is $N$.
  
Further, given a point-line incidence geometry $\Gamma(P, L)$, a {\it geometric hyperplane} of $\Gamma(P, L)$ is a subset of its point set such that
a line of the geometry is either fully contained in the subset or has with it just a single point in common; if all the lines passing through a given point lie in the hyperplane, the point in question is called deep.
The {\it Veldkamp space} of $\Gamma(P, L)$, $\mathcal{V}(\Gamma)$,  is the space \cite{buco} in  which a point is a geometric hyperplane of  $\Gamma$ and 
a line is the collection $H'H''$ of all geometric hyperplanes $H$ of $\Gamma$  such that $H' \cap H'' = H' \cap H = H'' \cap H$ or $H = H', H''$, where $H'$ and $H''$ are distinct points of $\mathcal{V}(\Gamma)$.
For a $\Gamma(P, L)$ with {\it three} points on a line, all Veldkamp lines are of the form
$\{H', H'', \overline{H' \Delta H''}\}$ where $\overline{H' \Delta H''}$
is the complement of symmetric difference of $H'$ and $H''$, i.\,e. they form a
vector space over $GF(2)$. In what follows we shall denote $\overline{H' \Delta H''}$ as $H' \oplus H''$ and call it the {\it Veldkamp sum}  of 
$H'$ and $H''$.

Finally, a subset of the point-set of $\Gamma(P, L)$ is called a {\it subspace} iff any line from $L$ intersects it in zero, one or all of its points. A subspace is called {\it singular} if any two of its points are collinear. If $p$ is a point, the symbol $p^{\perp}$ denotes the set of all points collinear with it, including the point itself. A point-line incidence structure $\Gamma(P, L)$ is called a {\it gamma} space iff $p^{\perp}$ is a subspace for every point $p \in P$.

\section{Veldkamp Space of the Doily}
As it is well known \cite{twoqub}, the doily features three different types of geometric hyperplanes (that is, Veldkamp points), namely six ovoids, fifteen perp-sets and ten grids. An ovoid  is a set of points such that each line is incident with exactly one point in it; hence, each ovoid of the doily has five points.
Using the duad-syntheme representation of the doily, it can be written as
\begin{equation}
o_i = \{\{i,j\}| j \in S \setminus \{i\}\}, \qquad i \in S.
\end{equation}  
A perp-set is the set of points collinear with a given point, the point inclusive. If the latter is $\{i,j\}$, and  taking $$\{1,2,3,4,5,6\} = \{i,j,k,l,m,n\}$$ in some order, the corresponding perp-set $p_{ij}$ reads
\begin{equation}
p_{ij} = \{\{i,j\}; \{k,l\}, \{k,m\}, \{k,n\}, \{l,m\}, \{l,n\}, \{m,n\} \}.
\end{equation}
Finally, the nine points of a grid, $g_{ijk}~ (= g_{lmn})$, can be represented as
\begin{equation}
g_{ijk} = \{\{a,b\}| a \in \{i,j,k\}~ {\rm and}~ b \in \{l,m,n\}\}.
\end{equation}
It can readily be verified that $p_{ij} = o_i \oplus o_j$ and $g_{ijk} = o_i \oplus o_j \oplus o_k$. The doily contains 155 Veldkamp lines that fall into five different families, having the following representatives
\begin{equation}
\{p_{ij}, g_{ikl}, g_{jkl}\}, 
\end{equation}
\begin{equation}
\{p_{ij}, p_{kl}, p_{mn}\}, 
\end{equation}
\begin{equation}
\{p_{ij}, p_{ik}, p_{jk}\}, 
\end{equation}
\begin{equation}
\{o_{i}, p_{jk}, g_{ijk}\}, 
\end{equation}
\begin{equation}
\{o_{i}, o_{j}, p_{ij}\}. 
\end{equation}

\section{Three Off-Doily Sectors of the Magic Veldkamp Line and the Doily's Veldkamp Space}
As already mentioned, a magic Veldkamp line of $W(5,2)$, the symplectic polar space behind the generalized three-qubit Pauli group, consists of a hyperbolic quadric $\mathcal{Q}^+(5,2)$, an elliptic quadric $\mathcal{Q}^-(5,2)$ and a quadratic cone $\widehat{\mathcal{Q}}(4,2)$ that have the doily ($ \cong \mathcal{Q}(4,2)$) in common \cite{lhs}. Our focus will be on the complements of the doily of the three constituents (to be called sectors in the sequel), in particular on how the points in these sectors are related with geometric hyperplanes of the core doily.

\subsection{Grids and Hyperbolic Sector}
It is easy to show that the 20 off-doily points of the hyperbolic sector form ten complementary pairs, the two points in any such pair being related to a particular grid of the doily.
Let us label these twenty points by three-element subsets of $S=\{1,2,3,4,5,6\}$ and let $\{a,b,c,i,j,k\}$ be a partition of $S$. Then the nine lines passing through the point $abc$
are\footnote{In what follows we will use a short-hand notation $\{a\} = a$,  $\{a,b\} = ab$, etc.} 
$$\{abc, aij, ak\}, \{abc, aik, aj\}, \{abc, ajk, ai\},$$
$$\{abc, bij, bk\}, \{abc, bik, bj\}, \{abc, bjk, bi\},$$
$$\{abc, cij, ck\}, \{abc, cik, cj\}, \{abc, cjk, ci\},$$
and the nine lines through the complementary/conjugate point $ijk$ read
$$\{ijk, iab, ic\}, \{ijk, iac, ib\}, \{ijk, ibc, ia\},$$
$$\{ijk, jab, jc\}, \{ijk, jac, jb\}, \{ijk, jbc, ja\},$$
$$\{ijk, kab, kc\}, \{ijk, kac, kb\}, \{ijk, kbc, ka\}.$$
Comparing these two sets of equations with eq.\,(3) we see that in both cases the lines cut the core doily in the nine points of the same {\it grid}, namely the $g_{abc}$ one. Figure 1 illustrates this property for two pairs of complementary points, 146/235 ({\it top}) and 136/245 ({\it bottom}).

\begin{figure}[pth!]
\centerline{\includegraphics[width=10.cm,height=10.cm,clip=]{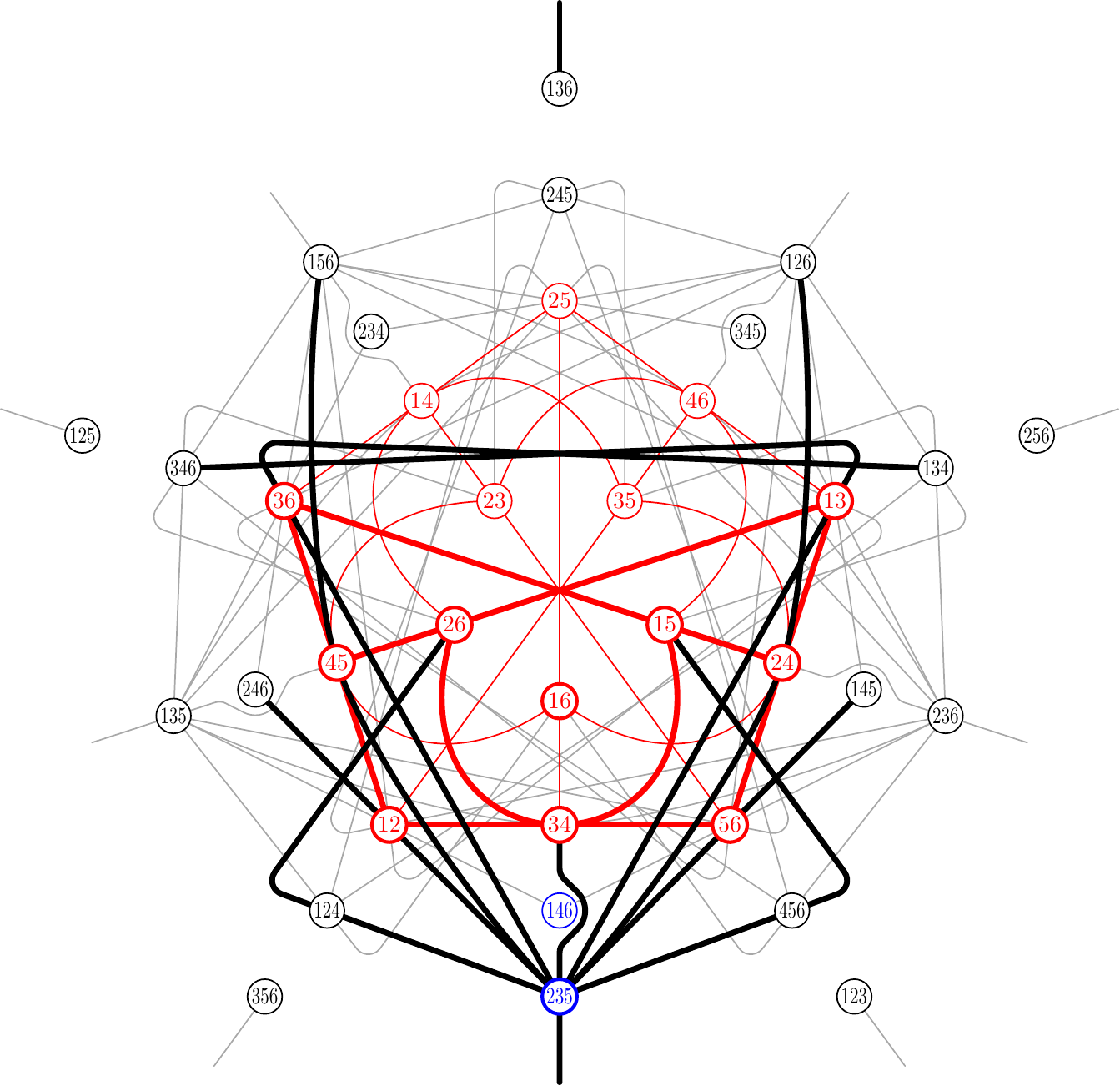}}
\vspace*{0.3cm}
\centerline{\includegraphics[width=10.cm,height=10.cm,clip=]{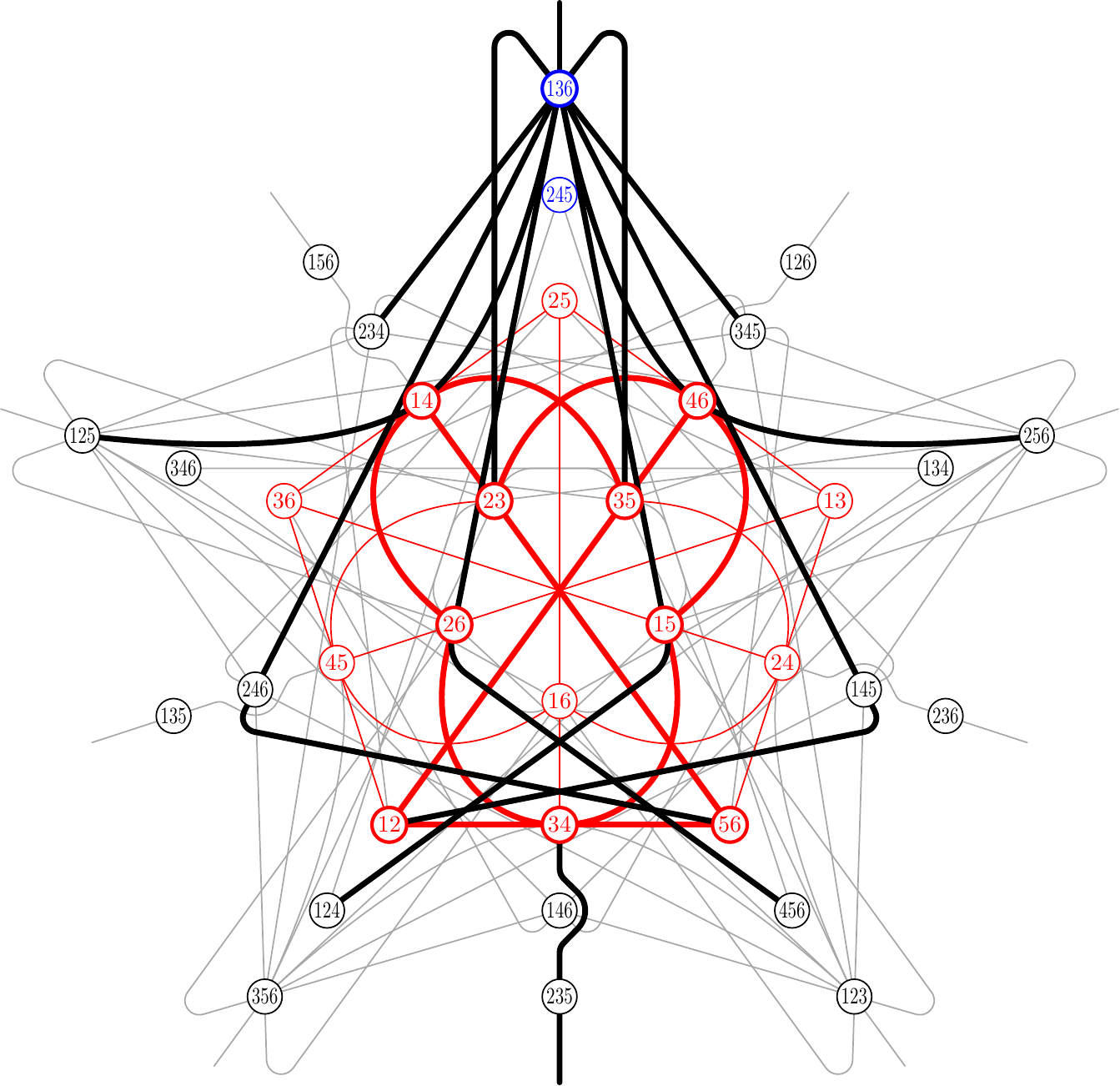}}
\caption{A model of $\mathcal{Q}^+(5,2)$ built around the doily (red); the nine lines (shown in bold) concurrent in an off-doily point (bold blue) cut the doily in a grid (bold red).}
\end{figure}

\subsection{Ovoids and Elliptic Sector}
The twelve off-doily points of the elliptic sector form six complementary pairs. If we label these points as $1,2,...,6$ and $1',2',...,6'$, and regard $i$ and $i'$ as complementary/conjugate, then the five lines
through an off-doily point $i$ or $i'$ are, respectively, of the form \cite{polster}
$$\{i,j', ij\}  \qquad {\rm or}  \qquad \{i',j, ij\},$$
where $j \in S$, $j' \in S'$, $i \neq j$ and $i' \neq j'$. Comparing these expressions with  eq.\,(1) we see that the five points of the doily in both cases correspond to the same {\it ovoid}, $o_i$. Figure 2 serves as an illustration of this fact for points 3 and 3'.
\begin{figure}[t]
\centerline{\includegraphics[width=11.cm,height=11.cm,clip=]{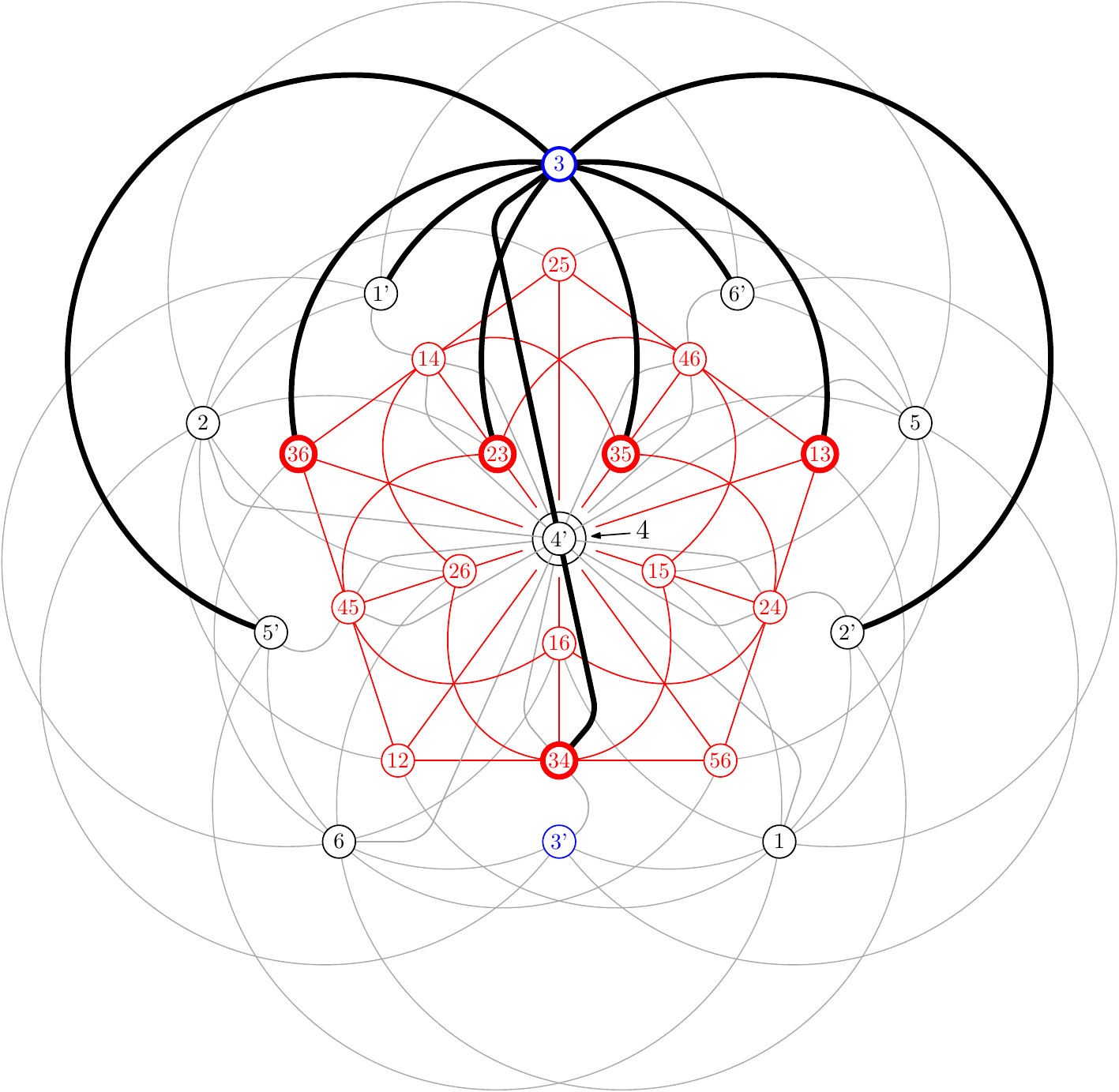}}
\caption{A model of $\mathcal{Q}^-(5,2)$ built around the doily (red); the five lines (shown in bold) concurrent in an off-doily point (bold blue) cut the doily in an ovoid (bold red).}
\end{figure}

\subsection{Perp-sets and Parabolic Sector}
The parabolic sector comprises  16 points of which one has a special footing, being the nucleus of $\mathcal{Q}(4,2)$, that is the point common to all tangent hyperplanes of the quadric. If we label this nucleus by $123456$ and the remaining 15 points by four-element subsets of $S$, $klmn$, then we can set up a natural bijection between these 15 points and the 15 perp-sets of the doily, each being defined by its unique deep point $ij$, in the form
$$\{123456, klmn, ij\}.$$  

\subsection{Sectors Image of the Doily's Veldkamp Space}
Based on these findings, we can establish the following correspondence between the geometric hyperplanes of the doily and off-doily points of the three sectors,
$$o_i       \mapsto {\rm a~pair~ of~complementary~points}~ i/i'~ {\rm in~ the~ elliptic~ sector}, $$
$$p_{ij}    \mapsto {\rm a~single~point}~klmn~ {\rm in~ the~ parabolic~ sector}, $$
$$g_{ijk}   \mapsto {\rm a~pair~ of~complementary~points}~ijk/lmn~ {\rm in~ the~ hyperbolic~ sector}, $$
from where we readily get the following sectorial counterparts of the five families of Veldkamp lines of the doily (eqs.\,(4)--(8), respectively)
\begin{equation}
\{klmn, ikl/jmn, jkl/imn\}, 
\end{equation}
\begin{equation}
\{klmn, ijmn, ijkl\}, 
\end{equation}
\begin{equation}
\{klmn, jlmn, ilmn\}, 
\end{equation}
\begin{equation}
\{i/i', ilmn, ijk/lmn\}, 
\end{equation}
\begin{equation}
\{i/i', j/j', klmn\}. 
\end{equation}
Note that the parabolic sector plays a (slightly) different role than the other two.

\section{Towards Parapolar Spaces}
There are several interesting implications of this correspondence. We will mention only one that we find particularly intriguing as it concerns
so-called {\it parapolar} spaces. A point-line incidence geometry $\Gamma(P, L)$ is called a parapolar space iff it satisfies the following 
properties \cite{shult}: (i)  $\Gamma$ is a connected gamma space, (ii) for every line $l \in L$, $l^{\perp}$ is not a singular subspace, and (iii)  
for every pair of non-collinear points $x,y \in P$, $x^{\perp} \cap y^{\perp}$ is either empty, a single point, or a non-degenerate polar space of rank at least two in which case $(x,y)$ is called a polar pair. And it is the very last property that makes a parapolar space to be a worth exploring concept in QIT. Indeed, as a grid of the doily, being isomorphic to $\mathcal{Q}^{+}(3,2)$, is, like the doily itself, a non-degenerate polar space of rank two, the corresponding pair of complementary points in the associated hyperbolic sector play in our model the role of polar pairs. Moreover, if we regard 
$W(5,2)$ as being embedded in some parapolar space like, for example, $L_3 \times W(5,2)$ where $L_3$ is a projective line of size three \cite{kasshu}, then such a pair of complementary points becomes
a polar pair in this parapolar space.\footnote{We note that the last requirement in axiom (iii) of the definition of a parapolar space puts aside pairs of complementary points in the {\it elliptic} sector, because an ovoid of the doily is isomorphic to  $\mathcal{Q}^{-}(3,2)$ and the latter is a non-degenerate polar space of rank {\it one} only (see Sec.\,2).}

The fact that the above-described  three-qubit Veldkamp line setting for the Veldkamp space of the doily leads rather straightforwardly to parapolar spaces should not come as a surprise. Parapolar spaces were originally invented to characaterize Lie geometries associated with exceptional algebraic groups \cite{cohco}, the groups that play a very important role in physics. Recently, they were also linked with the properties of  the Freudenthal-Tits Magic Square \cite{mald}, which encodes relation between a certain class of also physically-relevant semi-simple Lie algebras. In light of these facts, the eventual occurrence of parapolar spaces in the context of QIT is a rather viable conjecture.

\section*{Acknowledgments}
This work was supported by the Slovak Research and Development Agency under the contract $\#$ SK-FR-2017-0002, as well as by the Slovak VEGA Grant Agency, Project $\#$ 2/0003/16.

\end{document}